\documentclass{article}%
\usepackage{amsmath}%
\usepackage{amsfonts}%
\usepackage{amssymb}%
\usepackage{graphicx}

\begin{document}

\title{An Amplitude-Phase (Ermakov-Lewis) Approach for the Jackiw-Pi Model of Bilayer Graphene}
\author{{\tt Kira V. Khmelnytskaya} and {\tt Haret C. \ Rosu}\\Potosinian Institute of Science and Technology,\\Apdo Postal 3-74 Tangamanga, 78231 San Luis Potos\'{\i}, Mexico}

\maketitle

\medskip

\begin{center} J. Phys. A 42 (2009) 042004 (11 pp)\\
Fast Track Communication \end{center}

\begin{abstract}
\noindent In the context of bilayer graphene we use the simple gauge model of Jackiw and
Pi to construct its numerical solutions in powers of the bias potential $V$
according to a general scheme due to Kravchenko. Next, using this numerical
solutions, we develop the Ermakov-Lewis approach for the same model. This
leads us to numerical calculations of the Lewis-Riesenfeld phases that could
be of forthcoming experimental interest for bilayer graphene. We also present
a generalization of the Ioffe-Korsch nonlinear Darboux transformation.\\

\medskip


\end{abstract}

\section{\bigskip Introduction}

The recent discovery by Novoselov et al \cite{nov04} that a single atomic
layer of graphite usually known as graphene can be stable under some special
experimental conditions opened widely the realm of a soon-to-come technology
of graphene-based electronic devices. The main transport mechanism of the
charged carriers is in this case an enhanced quantum Hall effect accompanied
by a non-zero Berry's geometric phase that has been experimentally observed
\cite{yzhang05}. On the other hand, the quantum Hall effect and the Berry
phase have been also discussed for graphene bilayers \cite{mcC-F06} that
technologically could be even more interesting than graphene itself. Motivated
by these works, in this paper, we introduce the more general Lewis-Riesenfeld
phases for the case of a graphene bilayer. These phases are defined in a
mathematical formalism called the Ermakov-Lewis approach, which is shortly
presented. For the bilayer, we will make usage of the recent gauged Dirac
model of Jackiw and Pi \cite{JP2008}. With the aid of a mathematical procedure
due to Kravchenko \cite{KrCV2008} we present the results in a series expansion
in the bias potential $V$ applied to the bilayer graphene for any gauge
potential $\mathcal{A}$. We give detailed numerical results for a particular
case of $\mathcal{A}$. \bigskip

\section{\bigskip Jackiw-Pi model for bilayer graphene}

Recently Jackiw and Pi introduced a simple gauged Dirac model of bilayer
graphene that can be reduced to the following coupled system of first-order
differential equations \cite{JP2008}
\begin{align}
(D_{r}-\Phi(r))u(r)  & =Vv(r),\label{dir1}\\
(D_{r}+\Phi(r))v(r)  & =-Vu(r),\label{dir}%
\end{align}
where $D_{r}=\frac{d}{dr},\quad V$ is a constant representing the external
bias voltage and%
\begin{equation}
\Phi(r)=\frac{k/2}{r}-\frac{k\mathcal{A}(r)}{r}~.\label{Fi}%
\end{equation}
Physically, function $\Phi$ has two components:

\noindent$\bullet$ a scalar field $\varphi=\frac{k/2}{r}$ that describes a
particular dimerization called Kekul\'{e} distortion \cite{Hou07} which can be
also interpreted as characterizing the condensate arising from states bound by
interlayer Coulomb forces between particles in one layer and holes in the
other; the fingerprint of $\varphi$ in $\Phi$ is the parameter $k$ which
should be an odd integer number.

\noindent$\bullet$ a vector potential, $\mathcal{A}$, which was first
introduced by Jackiw and Pi for a monolayer graphene \cite{JP2007} to unpin
the vortices and then studied in the context of bilayer graphene as well
\cite{JP2008}. $\mathcal{A}$ is a gauge field satisfying some appropriate
conditions in the origin and at infinity. In \cite{JP2008} it was supposed
that
\begin{equation}
\mathcal{A}(0)=0\text{ and}\;\mathcal{A}(\infty)=\frac{1}{2}.\label{A}%
\end{equation}

It is easy to see that the spinor components $u$ and $v$ are necessarily
solutions of the following second-order differential equations obtained
directly from system (\ref{dir1})-(\ref{dir})%

\begin{align}
-\left(  D_{r}+\Phi(r)\right)  \left(  D_{r}-\Phi(r)\right)  u  &
=V^{2}u,\label{second1}\\
-\left(  D_{r}-\Phi(r)\right)  \left(  D_{r}+\Phi(r)\right)  v  &
=V^{2}v.\label{second2}%
\end{align}
This couple of equations can be rewritten in the following way%
\begin{align}
-D_{r}^{2}u+q_{1}(r,V)u  & =0,\;\text{where}\;q_{1}(r,V)=\Phi^{2}+\Phi
^{\prime}-V^{2},\label{schred1}\\
-D_{r}^{2}v+q_{2}(r,V)v  & =0,\;\text{where}\;q_{2}(r,V)=\Phi^{2}-\Phi
^{\prime}-V^{2}.\label{schred2}%
\end{align}
The latter form of the uncoupled system shows that the two functions $u$ and
$v$ in the Jackiw-Pi model are supersymmetric partners in supersymmetric
quantum mechanics and therefore many known supersymmetric results can be
directly applied.

\section{Kravchenko's representation for the general Schr\"{o}dinger solution}

Consider the Sturm-Liouville equation%

\begin{equation}
D_{r}(pD_{r}u)+qu=\omega^{2}u,\label{sch}%
\end{equation}
where $p$ and $q$ are complex-valued functions of a real variable $r\in\lbrack
a,b],\quad p\in C^{1}(a,b)$ is bounded and non-vanishing on $[a,b]$ and
$\omega$ is an arbitrary complex constant. Suppose that there exists a
solution $g_{0}$ of the equation
\[
D_{r}(pD_{r}g_{0})+qg_{0}=0
\]
on $(a,b)$ such that $g_{0}\in C^{2}(a,b)$ together with $\frac{1}{g_{0}}$ are
bounded on $[a,b].$

Kravchenko \cite{KrCV2008} proved that the general solution of (\ref{sch}) has
the form%
\begin{equation}
u=c_{1}u_{1}+c_{2}u_{2}\label{sol}%
\end{equation}
where $c_{1}$, $c_{2}$ are arbitrary complex constants and%
\begin{equation}
u_{1}=g_{0}%
{\displaystyle\sum\limits_{\text{even}\hspace{0.05in}n=0}^{\infty}}
\frac{\omega^{^{n}}}{n!}\widetilde{X}^{(n)}\,,\,\qquad u_{2}=g_{0}%
{\displaystyle\sum\limits_{\text{odd}\hspace{0.05in}n=1}^{\infty}}
\frac{\omega^{^{n-1}}}{n!}X^{(n)}\label{u1u2}%
\end{equation}
with $\widetilde{X}^{(n)}$ and $X^{(n)}$ being defined by the following
recursive relations%
\[
\widetilde{X}^{(0)}\equiv1,\qquad X^{(0)}\equiv1,
\]%
\begin{equation}
\widetilde{X}^{(n)}(r)=\left\{
\begin{tabular}
[c]{ll}%
$n\int_{a}^{r}\widetilde{X}^{(n-1)}(\xi)g_{0}^{2}(\xi)d\xi\qquad$ & $\text{for
an odd }n$\\
& \\
$n\int_{a}^{r}\widetilde{X}^{(n-1)}(\xi)\frac{d\xi}{p(\xi)g_{0}^{2}(\xi
)}\qquad$ & $\text{for an even }n$%
\end{tabular}
\right. \label{Xtil}%
\end{equation}%
\begin{equation}
X^{(n)}(r)=\left\{
\begin{tabular}
[c]{ll}%
$n\int_{a}^{r}X^{(n-1)}(\xi)\frac{d\xi}{p(\xi)g_{0}^{2}(\xi)}\qquad$ &
$\text{for an odd }n$\\
& \\
$n\int_{a}^{r}X^{(n-1)}(\xi)g_{0}^{2}(\xi)d\xi\qquad$ & $\text{for an even
}n.$%
\end{tabular}
\right. \label{X}%
\end{equation}
This iterative scheme is appropriate for an easy implementation of numerical
solutions and therefore for not exactly-solvable problems.

\section{Application to the Jackiw-Pi uncoupled system}

To apply Kravchenko's procedure to the Jackiw-Pi uncoupled system we start
with the equation for $u$. It is easy to find a particular solution
$g_{0\text{ }}$ from the factorized form of the $u$ equation (\ref{second1}).
We immediately get
\[
\frac{g_{0}\prime}{g_{0}}=\frac{k}{r}\left(  \frac{1}{2}-A(r)\right)
,\quad\longrightarrow g_{0}(r)=r^{\frac{k}{2}}e^{-k%
{\displaystyle\int}
\frac{A(r)}{r}dr},
\]
and the general solution $u$ has the form $u=C_{1}u_{1}+C_{2}u_{2}$ , with
\begin{equation}
u_{1}=\frac{g_{0}(r)}{g_{0}(a)}%
{\displaystyle\sum\limits_{\text{even}\hspace{0.05in}n=0}^{\infty}}
\frac{\widetilde{X}^{(n)}}{n!}V^{^{n}}\,,\qquad\ u_{2}=-g_{0}(a)g_{0}(r)%
{\displaystyle\sum\limits_{\text{odd}\hspace{0.05in}n=1}^{\infty}}
\frac{X^{(n)}}{n!}V^{^{n-1}},\label{u}%
\end{equation}
where $\widetilde{X}^{(n)}$ and $X^{(n)}$ are as in (\ref{Xtil}) and
(\ref{X}), respectively, for $p\equiv-1$.

The general solution $v(r)$ of equation (\ref{second2}) can be obtained now by
means of the Darboux transformation $v=C_{1}v_{1}+C_{2}v_{2}=\frac{1}{V}%
g_{0}D_{r}g_{0}^{-1}(C_{1}u_{1}+C_{2}u_{2})$%

\begin{equation}
v_{1}=-\frac{1}{g_{0}(a)g_{0}(r)}%
{\displaystyle\sum\limits_{\text{odd}\hspace{0.05in}n=1}^{\infty}}
\frac{\widetilde{X}^{(n)}}{n!}V^{^{n}}\,,\,\qquad v_{2}=\frac{g_{0}(a)}%
{Vg_{0}(r)}%
{\displaystyle\sum\limits_{\text{even}\hspace{0.05in}n=0}^{\infty}}
\frac{X^{(n)}}{n!}V^{^{n}}.\label{v}%
\end{equation}
By construction, $u_{1}$ and $u_{2\text{ }}$ satisfy the following initial
conditions
\begin{equation}
u_{1}(a)=1,\ u_{1}^{\prime}(a)=\Phi(a),\ u_{2}(a)=0,\ u_{2}^{\prime
}(a)=1,\label{condu1u2}%
\end{equation}
with their Wronskian being equal to $W=1.$ Consequently, the functions $v_{1}
$ and $v_{2}$ take the initial values
\begin{equation}
v_{1}(a)=0,\quad v_{1}^{\prime}(a)=-V,\quad v_{2}(a)=\frac{1}{V},\quad
v_{2}^{\prime}(a)=-\frac{\Phi(a)}{V}.\label{condv1v2}%
\end{equation}

If one takes $V=0$ in (\ref{second1}), the formulas (\ref{u}) give
\[
u_{1}\left(  r\right)  =c_{1}g_{0}\left(  r\right)  \text{ and }u_{2}\left(
r\right)  =c_{2}g_{0}\left(  r\right)  \int_{a}^{r}\frac{d\xi}{g_{0}^{2}(\xi
)},
\]
where the expression in $u_{2}$ represents the well known construction of a
second linearly independent solution of a Schr\"{o}dinger equation.

When $r\rightarrow\infty$, $\mathcal{A\rightarrow}\frac{1}{2}$, the system
(\ref{dir1})-(\ref{dir}) reads as%
\[
D_{r}u-Vv=0,\qquad D_{r}v+Vu=0.
\]
A solution $g_{0}$ in this case is an arbitrary constant, e.g., $g_{0}\equiv
1$. Calculating $\widetilde{X}^{(n)}$ and $X^{(n)}$ with the aid of
(\ref{Xtil})-(\ref{X}) we obtain%

\[
\widetilde{X}^{(n)}(r)=\left\{
\begin{tabular}
[c]{ll}%
$(-1)^{\frac{n-1}{2}}r^{n},$ & $\text{for an odd }n$\\
& \\
$(-1)^{\frac{n}{2}}r^{n},$ & $\text{for an even }n$%
\end{tabular}
\right.  ,\;X^{(n)}(r)=\left\{
\begin{tabular}
[c]{ll}%
$(-1)^{\frac{n+1}{2}}r^{n},$ & $\text{for an odd }n$\\
& \\
$(-1)^{\frac{n}{2}}r^{n},$ & $\text{for an even }n.$%
\end{tabular}
\right.
\]
Now substitution into (\ref{u}) and (\ref{v}) gives us the following solutions
(cf \cite{JP2008})
\[
u_{1}=\cos Vr,\;u_{2}=\frac{1}{V}\sin Vr,\;v_{1}=-\sin Vr,\;v_{2}=\frac{1}%
{V}\cos Vr.
\]

When $r\rightarrow0,$ and consequently $\mathcal{A}\rightarrow0$, the
superpotential $\Phi$ takes the form $\Phi=\frac{k}{2r}$. For system
(\ref{dir1}), (\ref{dir}) with this particular potential, in \cite{JP2008} the
following solution is presented%
\begin{equation}
u=\sqrt{r}J_{(k-1)/2}(Vr),\qquad v=-\sqrt{r}J_{(k+1)/2}(Vr),\label{SolJP}%
\end{equation}
which is easily obtained from (\ref{u}) and (\ref{v}) choosing $g_{0}%
=r^{\frac{k}{2}}$ and calculating $u_{1}$ and $v_{1}.$ Indeed, for
$\widetilde{X}^{(n)}$ we obtain the formulas%

\[
\widetilde{X}^{(n)}(r)=\left\{
\begin{tabular}
[c]{ll}%
$(-1)^{\frac{n-1}{2}}\frac{n!!(k-1)!!}{(k+n)!!}r^{k+n},\qquad$ & $\text{for an
odd }n$\\
& \\
$(-1)^{\frac{n}{2}}\frac{(n-1)!!(k-1)!!}{(k+n-1)!!}r^{n},\qquad$ & $\text{for
an even }n$%
\end{tabular}
\right.
\]
leading us to the solution (\ref{SolJP}) multiplied by a constant.

We have now all the elements to proceed to specific calculations. Let us
consider the Dirac system (\ref{dir1})-(\ref{dir}) choosing in (\ref{Fi}) the
vector potential $\ \mathcal{A}(r)=\frac{1}{2}-\frac{1}{2}e^{-\mu r}$ which is
in agreement with conditions (\ref{A}). The parameter $\mu^{-1}$ plays the
role of a screening length. The superpotential (\ref{Fi}) then becomes
$\Phi(r)=\frac{k/2}{r}e^{-\mu r}$ and in this case the system (\ref{schred1}%
)-(\ref{schred2}) has the following form%
\begin{align}
-D_{r}^{2}u+\frac{k}{2r}e^{-\mu r}(\frac{k}{2r}e^{-\mu r}-\frac{1}{r}-\mu)u  &
=V^{2}u,\label{scherd_sys}\\
-D_{r}^{2}v+\frac{k}{2r}e^{-\mu r}(\frac{k}{2r}e^{-\mu r}+\frac{1}{r}+\mu)v  &
=V^{2}v,\label{scherd_sys1}%
\end{align}
The plots of $u$ and $v$ calculated within Kravchenko's scheme are presented
in Fig.1.
\begin{center}
\includegraphics[
height=2.7622in,
width=5.0055in
]%
{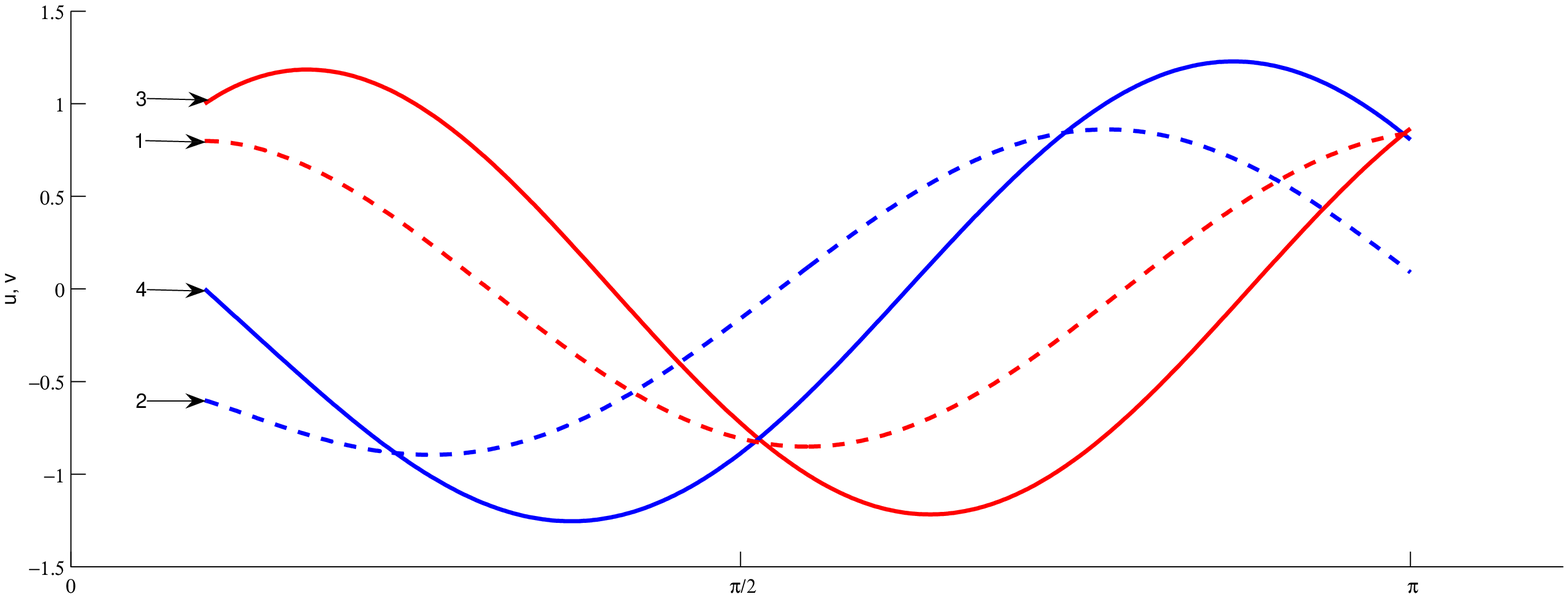}%
\\
Fig.1. The solutions $u(r)$ (red) and $v(r)$ (blue) of the supersymmetric
system (\ref{scherd_sys})-(\ref{scherd_sys1}) for $k=1$ and two cases of the
arbitrary constants $C_{1}$ and $C_{2}$ : --- is for $C_{1}=1,C_{2}=0$, and -
- - is for $C_{1}=\frac{1}{\sqrt{\Phi(a)}},C_{2}=-\Phi(a)$.
\end{center}

\bigskip

\section{The Ermakov-Lewis approach}

We will now expound on the Ermakov-Lewis (EL) formalism for the supersymmetric
system (\ref{schred1})-(\ref{schred2}). \bigskip If the functions $u$ and $v$
are written in the form \cite{Milne}%
\begin{equation}
u(r)=\alpha\rho_{u}(r)\sin(\varphi_{u}(r,V)-\beta),\;v(r)=\gamma\rho
_{v}(r)\sin(\varphi_{v}(r,V)-\delta),\label{y1}%
\end{equation}
where $\alpha,\beta,\gamma$and $\delta$ are arbitrary constants, then both the
amplitude functions $\rho_{u}$ and $\rho_{v}$ and the phase functions
$\varphi_{u}$ and $\varphi_{v}$ are directly related to the
Ermakov-Milne-Pinney nonlinear equations \cite{Pinney} $\ $%
\begin{equation}
\rho_{u,v}^{\prime\prime}(r)+q_{1,2}(r,V)\rho_{u,v}(r)=\frac{\lambda}%
{\rho_{u,v}^{3}(r)},\;\label{ermakov}%
\end{equation}
where $\lambda$ is an arbitrary constant, in the following way. While
$\rho_{u}$ and $\rho_{v}$ are solutions of (\ref{ermakov}) with a respective
potential, the phase functions $\varphi_{u}$ and $\varphi_{v}$ are expressed
in terms of $\rho_{u}$ and $\rho_{v}$ as follows%
\begin{equation}
\varphi_{u,v}(r,V)=%
{\displaystyle\int\limits_{a}^{r}}
\frac{1}{\rho_{u,v}^{2}(r^{\prime})}dr^{\prime}.\label{phase}%
\end{equation}

For $\rho_{u}$ and $\rho_{v}$ there is an elegant representation
\cite{EliezGr} in terms of pairs of linearly independent solutions
$(u_{1},u_{2})$ and $(v_{1},v_{2})$ of (\ref{schred1}) and (\ref{schred2})
respectively. Namely, general solutions of (\ref{ermakov}) can be written in
the following way
\begin{equation}
\rho_{u}=\sqrt{Au_{1}^{2}+Bu_{2}^{2}+2Cu_{1}u_{2}},\;\rho_{v}=\sqrt{Av_{1}%
^{2}+Bv_{2}^{2}+2Cv_{1}v_{2}},\label{ro}%
\end{equation}
where $A,$ $B$ and $C$ are constants related by the equality $AB-C^{2}%
=\frac{\lambda}{W^{2}}$, with $W$ being the Wronskian $u_{1}u_{2}^{\prime
}-u_{2}u_{1}^{\prime}=v_{1}v_{2}^{\prime}-v_{2}v_{1}^{\prime}$ (which is a
constant that does not change through the Darboux transformation).

The solution of the Ermakov equations allows us to obtain another important
dynamic invariant quantity $I_{u}$ and $I_{v}$ respectively, called the Lewis
invariant \cite{Lewis}. $I_{u}$ is constructed in terms of any solution $u$ of
(\ref{schred1}) and the corresponding solution $\rho_{u}$ of the
Ermakov-Milne-Pinney nonlinear equation (\ref{ermakov}) (and analogously
$I_{v}$ is constructed in terms of $v$ and $\rho_{v}$)%

\begin{equation}
I_{u}=\frac{1}{2}\left(  \frac{\lambda u^{2}}{\rho_{u}^{2}}+\left(  \rho
_{u}u^{\prime}-\rho_{u}^{\prime}u\right)  ^{2}\right)  ,\;I_{v}=\frac{1}%
{2}\left(  \frac{\lambda v^{2}}{\rho_{v}^{2}}+\left(  \rho_{v}v^{\prime}%
-\rho_{v}^{\prime}v\right)  ^{2}\right)  .\label{inv}%
\end{equation}
$\bigskip$

For the fixed interval $[a,b]$ the expressions $\frac{1}{\pi}\varphi
_{u,v}(b,E)$ give us the quantum number function depending on the energy $E$
\cite{Milne},\cite{Korsch85},\cite{IoffeKorsch}
\begin{equation}
N(E)=\frac{1}{\pi}%
{\displaystyle\int\limits_{a}^{b}}
\frac{1}{\rho^{2}(r)}dr,\label{qnumber}%
\end{equation}
used in standard quantum mechanics to identify bound states since at a bound
state energy $E_{n}$ the function $N(E)$\ takes integer values $N(E_{n}%
)=n+1,\quad n=0,1,...$ and this does not depend on the choice of the constants
$A,B$, and $C$.\ However, if the energy is different from $E_{n}$ then
(\ref{qnumber}) is not uniquely determined, i.e., for $E_{n}<E<E_{n+1}$ the
function $N(E)$ depends on the boundary conditions imposed to $\rho$.

An important point of the EL formalism is that the phase function
(\ref{phase}) also known as Lewis-Riesenfeld phase \cite{LR}, has two
components: the usual dynamic phase and a geometric phase that are given by
the following formulas (\cite{mor})%

\begin{equation}
\Delta\varphi_{u,v}^{dyn}(r)=%
{\displaystyle\int\limits_{a}^{r}}
\left[  \frac{1}{\rho_{u,v}^{2}(r^{\prime})}-\frac{1}{2}\frac{d}{dr^{\prime}%
}\left(  \rho_{u,v}(r^{\prime})\frac{d\rho_{u,v}(r^{\prime})}{dr^{\prime}%
}\right)  +\left(  \frac{d\rho_{u,v}(r^{\prime})}{dr^{\prime}}\right)
^{2}\right]  dr^{\prime}\label{dynamic}%
\end{equation}
and%
\begin{equation}
\Delta\varphi_{u,v}^{geom}(r)=%
{\displaystyle\int\limits_{a}^{r}}
\left[  \frac{1}{2}\frac{d}{dr^{\prime}}\left(  \rho_{u,v}(r^{\prime}%
)\frac{d\rho_{u,v}(r^{\prime})}{dr^{\prime}}\right)  -\left(  \frac
{d\rho_{u,v}(r^{\prime})}{dr^{\prime}}\right)  ^{2}\right]  dr^{\prime
},\label{geom}%
\end{equation}
Any of these phases can be calculated by employing Kravchenko's iterative
series (\ref{u}) and (\ref{v}) for the $\rho$ functions (\ref{ro}). Denoting%
\begin{align*}
\widetilde{\widetilde{Z}}_{u}  & =%
{\displaystyle\sum\limits_{even\hspace{0.05in}n=0}^{\infty}}
{\displaystyle\sum\limits_{even\hspace{0.05in}i=0}^{n}}
\frac{\widetilde{X}^{(i)}\widetilde{X}^{(n-i)}}{i!(n-i)!}V^{^{n}},\;Z_{u}=%
{\displaystyle\sum\limits_{odd\hspace{0.05in}n=1}^{\infty}}
{\displaystyle\sum\limits_{odd\hspace{0.05in}i=1}^{n}}
\frac{X^{(i)}X^{(n-i)}}{i!(n-i)!}V^{^{n-1}},\\
\;\widetilde{Z}_{u}  & =%
{\displaystyle\sum\limits_{odd\hspace{0.05in}n=1}^{\infty}}
{\displaystyle\sum\limits_{even\hspace{0.05in}i=0}^{n-1}}
\frac{\widetilde{X}^{(i)}X^{(n-i)}}{i!(n-i)!}V^{^{n}}%
\end{align*}
we get%

\begin{equation}
\rho_{u}(r)=g_{0}(r)\left[  \frac{A}{g_{0}^{2}(a)}\widetilde{\widetilde{Z}%
}_{u}\,+Bg_{0}^{2}(a)Z_{u}-\frac{2C}{V}\widetilde{Z}_{u}\right]  ^{\frac{1}%
{2}}.\label{rou}%
\end{equation}
Introducing the notations%
\begin{align*}
\widetilde{\widetilde{Z}}_{v}  & =%
{\displaystyle\sum\limits_{odd\hspace{0.05in}n=1}^{\infty}}
{\displaystyle\sum\limits_{odd\hspace{0.05in}i=1}^{n}}
\frac{\widetilde{X}^{(i)}\widetilde{X}^{(n-i)}}{i!(n-i)!}V^{^{n+1}},\;Z_{v}=%
{\displaystyle\sum\limits_{even\hspace{0.05in}n=0}^{\infty}}
{\displaystyle\sum\limits_{even\hspace{0.05in}i=0}^{n}}
\frac{X^{(i)}X^{(n-i)}}{i!(n-i)!}V^{^{n}},\\
\;\widetilde{Z}_{v}  & =%
{\displaystyle\sum\limits_{odd\hspace{0.05in}n=1}^{\infty}}
{\displaystyle\sum\limits_{even\hspace{0.05in}i=0}^{n-1}}
\frac{X^{(i)}\widetilde{X}^{(n-i)}}{i!(n-i)!}V^{^{n}},
\end{align*}
the amplitude function $\rho_{v}$ is calculated as follows%
\begin{equation}
\rho_{v}(r)=\frac{1}{g_{0}(r)}\left[  \frac{A}{g_{0}^{2}(a)}\widetilde
{\widetilde{Z}}_{v}+\frac{Bg_{0}^{2}(a)}{V}Z_{v}-\frac{2C}{V}\widetilde{Z}%
_{v}\right]  ^{\frac{1}{2}}\label{rov}%
\end{equation}
The particular values of the functions $\rho_{u}$ and $\rho_{v}$ can be
obtained by relating the constants $A,B$ and $C$ to (\ref{condu1u2}) and
(\ref{condv1v2}), that is%
\[
\rho_{u}(a)=\sqrt{A},\quad\rho_{u}^{\prime}(a)=\frac{A\Phi(a)+C}{\sqrt{A}},
\]%
\begin{equation}
\rho_{v}(a)=\frac{\sqrt{B}}{V},\quad\rho_{v}^{\prime}(a)=-\frac{\sqrt{B}%
\Phi(a)}{V}-\frac{VC}{\sqrt{B}}.\label{condrov}%
\end{equation}
Based on (\ref{rou}) and (\ref{rov}), we have calculated the Lewis-Riesenfeld
angular integrals (\ref{dynamic}) and (\ref{geom}) for the same decaying
exponential profile of $\mathcal{A}(r)$.The results for the geometric
Lewis-Riesenfeld phases are plotted in Fig. 2 and Fig. 3 of the paper for
$\lambda$ $=1$ and $\lambda=4$, respectively, and for the same finite interval.

One can check by direct calculation that the values of the Ermakov-Lewis
invariant (\ref{inv}) do not change under tthe supersymmetric transformation,
that is $I_{u}=I_{v}$. Indeed, noticing that $I_{u}$ and $I_{v}$ are
$r$-independent quantities, we can calculate them at one point only, e.g., at
$r=a$ and with arbitrary constants $C_{1},C_{2}$ and $A,B,C$. We have
\begin{align*}
I_{u}(a)  & =\frac{1}{2}\left(  AC_{2}^{2}+BC_{1}^{2}-2CC_{1}C_{2}\right)  ,\\
I_{v}(a)  & =\frac{1}{2}\left(  \frac{\lambda C_{2}^{2}}{B}+BC_{1}^{2}%
-2CC_{1}C_{2}+\frac{C^{2}C_{2}^{2}}{B}\right)  .
\end{align*}
Taking into account that $C^{2}=AB-\lambda$ we obtain $I_{u}(a)=I_{v}(a)$.

\section{Generalization of the Ioffe-Korsch intertwining formulas}

If the functions $u$ and $v$ are solutions of a coupled first-order system of
the Dirac type then the corresponding amplitude functions $\rho_{u}$ and
$\rho_{v}$ are related through nonlinear Dirac-Ermakov relationships that in a
particular case $\lambda=1$ have been obtained previously by Ioffe and Korsch
\cite{IoffeKorsch}. Consider the equality%
\[
\rho_{u}^{2}=Au_{1}^{2}+Bu_{2}^{2}+2Cu_{1}u_{2}.
\]
Substitution of $u_{1}=-\frac{1}{V}(D_{r}+\Phi)v_{1}\ $and $u_{2}=-\frac{1}%
{V}(D_{r}+\Phi)v_{2}$ leads to the relationship%
\[
V^{2}\rho_{u}^{2}=\Phi^{2}\rho_{v}^{2}+\Phi(\rho_{v}^{2})^{\prime}+S_{v},
\]
where $S_{v}=A\left(  v_{1}^{\prime}\right)  ^{2}+B\left(  v_{2}^{\prime
}\right)  ^{2}+2Cv_{1}^{\prime}v_{2}^{\prime}$. \ Taking into account
(\ref{y1}), the functions $v_{1}$ and $v_{2}$ can be expressed as
$v_{1}=\gamma_{1}\rho_{v}\sin(\varphi_{v}-\delta_{1})$ and $v_{2}=\gamma
_{2}\rho_{v}\sin(\varphi_{v}-\delta_{2})$. Calculating the constants
$\gamma_{i}$ and $\delta_{i}$ according to (\ref{condv1v2}) and (\ref{condrov}%
) we get
\begin{align*}
v_{1}  & =-\sqrt{B}\rho_{v}\sin\varphi_{v},\\
v_{2}  & =\frac{\rho_{v}}{\sqrt{B}}\left(  C\sin\varphi_{v}+\cos\varphi
_{v}\right)  .
\end{align*}
Now taking the derivatives and substituting them into the expression for
$S_{v}$ we obtain
\[
S_{v}=\left(  \rho_{v}^{\prime}\right)  ^{2}+\frac{1}{\rho_{v}^{2}}%
+(\lambda-1)\left(  \rho_{v}^{\prime}\sin\varphi_{v}+\frac{1}{\rho_{v}}%
\cos\varphi_{v}\right)  ^{2},
\]
where one can notice that $\left(  \rho_{v}^{\prime}\sin\varphi_{v}+\frac
{1}{\rho_{v}}\cos\varphi_{v}\right)  ^{2}=\frac{1}{B}\left(  v_{1}^{\prime
}\right)  ^{2}$.

Employing the same procedure with respect to $\rho_{u}^{2}$ and taking into
consideration that
\begin{align*}
u_{1}  & =\frac{\rho_{u}}{\sqrt{A}}\left(  -C\sin\varphi_{u}+\cos\varphi
_{u}\right)  ,\\
u_{2}  & =\sqrt{A}\rho_{u}\sin\varphi_{u},
\end{align*}
we arrive at the nonlinear Dirac-Ermakov system involving the amplitudes
$\rho_{u}$ and $\rho_{v}$\ and the phases $\varphi_{u},\varphi_{v}$ of the
Dirac spinors $u$ and $v$%
\begin{equation}
V^{2}\rho_{u}^{2}=\left(  (D_{r}+\Phi)\rho_{v}\right)  ^{2}+\frac{1}{\rho
_{v}^{2}}+(\lambda-1)\left(  \rho_{v}^{\prime}\sin\varphi_{v}+\frac{1}%
{\rho_{v}}\cos\varphi_{v}\right)  ^{2},\label{rou2}%
\end{equation}%
\begin{equation}
V^{2}\rho_{v}^{2}=\left(  (D_{r}-\Phi)\rho_{u}\right)  ^{2}+\frac{1}{\rho
_{u}^{2}}+(\lambda-1)\left(  \rho_{u}^{\prime}\sin\varphi_{u}+\frac{1}%
{\rho_{u}}\cos\varphi_{u}\right)  ^{2},\label{rov2}%
\end{equation}
where again it is worth noticing that $\left(  \rho_{u}^{\prime}\sin
\varphi_{u}+\frac{1}{\rho_{u}}\cos\varphi_{u}\right)  ^{2}=\frac{1}{A}\left(
u_{2}^{\prime}\right)  ^{2}$. Thus, there is a cross contribution to the
squares of the intertwined amplitude functions of one-unit jumps in $\lambda$
which is given by the squares of the derivatives of the corresponding
eigenfunctions. For $\lambda=1$ the system (\ref{rou2})-(\ref{rov2}) reduces
to the result in \cite{IoffeKorsch}.

\section{Application to the Jackiw-Pi model for screened-exponential form of
$\mathcal{A}(r)$}

To associate the spinor components $u$ and $v$ to the amplitude functions
$\rho_{u}$ and \ $\rho_{v}$ the relationship between the constants $C_{1}$ and
$C_{2}$ and $A,B$ and $C$ should be fixed. For instance, choosing $\rho
_{u}(a)=u(a)$, $\rho_{u}^{\prime}(a)=u^{\prime}(a)$ we get $A=C_{1}^{2}$,
$B=\frac{\lambda}{C_{1}^{2}}+C_{2}^{2}$, $C=C_{1}C_{2}$.

The corresponding $\Delta\varphi_{u,v}^{tot}=(n+1/2)\varphi_{u,v}$ and
$\Delta\varphi_{u,v}^{geom}$ \ for $\lambda=1$ and $\lambda=4$ are displayed
in Fig.2, Fig.3. and Fig.4. In addition, the amplitude functions for the same
values of the parameters are shown in Fig.5. Notice that since the problem is
undefined at the origin because of the $\frac{1}{r}$ singularity of $\Phi(r)$
the plots are displayed for the interval $[0.1\pi,\pi]$. For all plots we use
the following three sets of values of the constants $A,$ $B$ and $C$:

1. $A=1,$ $B=\lambda,$ $C=0$. These values imply $C_{1}=1,$ $C_{2}=0,$

$\quad u(r)=u_{1},\quad v(r)=v_{1}$.

2. $A=\frac{1}{\Phi(a)},$ $B=(\lambda+1)\Phi(a),$ $C=-1$. In this case
$C_{1}=\frac{1}{\sqrt{\Phi(a)}},$

$C_{2}=-\Phi(a),$ $\rho_{u}^{\prime}(a)=0$.

3. $A=\frac{\lambda}{V^{2}},$ $B=V^{2},$ $C=0$ $\ $leading to $C_{1}%
=\frac{\sqrt{\lambda}}{\Phi(a)},$ $C_{2}=0$.

\noindent The values of $V,\mu$ and $k$ are chosen as follows: $V=2.08236$, $\mu
=\frac{1}{20}$ and $k=1$.%

{\parbox[b]{2.8694in}{\begin{center}
\includegraphics[
height=2.8184in,
width=2.8694in
]%
{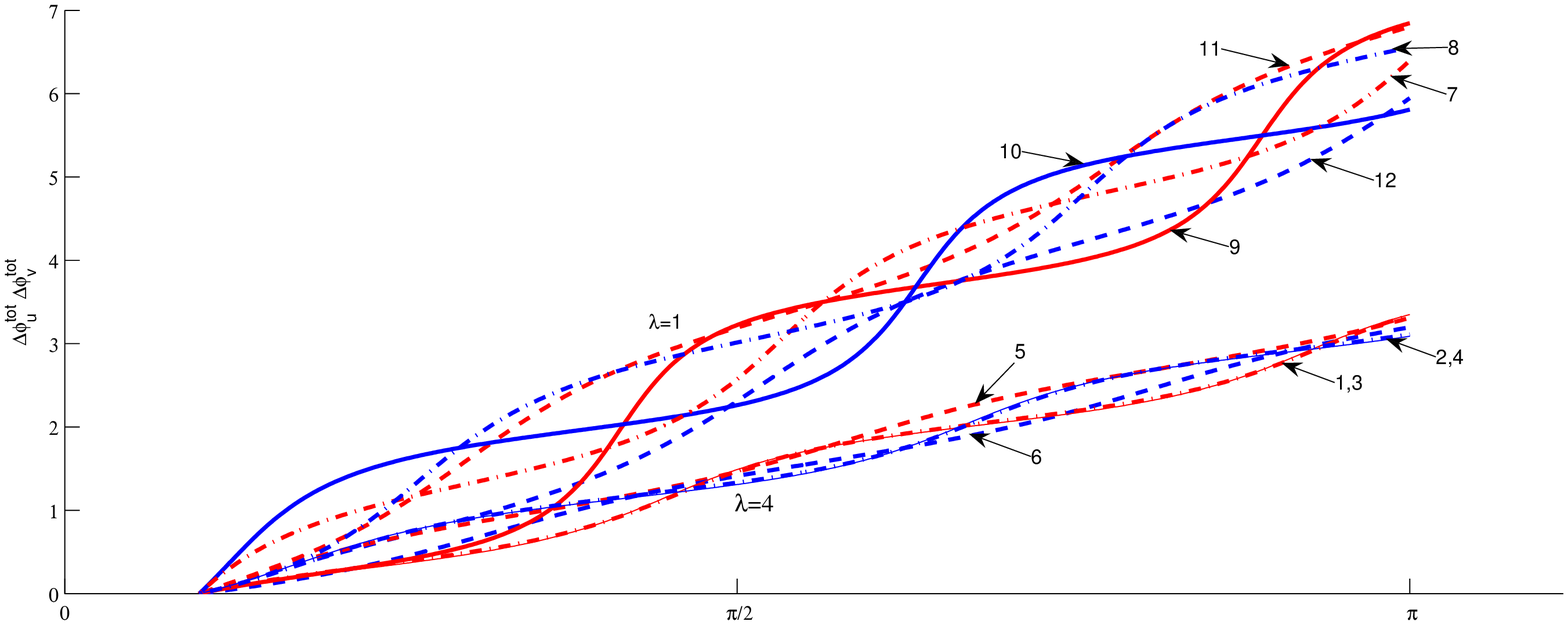}%
\\
Fig.2. $\Delta\varphi_{u}^{tot}$ (red) and $\ \Delta\varphi_{v}^{tot}$ (blue)
are plotted for three cases of the parameters $\ \ A,B$ and $C$ : --- is for
$A=1,B=\lambda,C=0$, - - - is for $A=\frac{1}{\Phi(a)},B=(\lambda
+1)\Phi(a),C=-1$, and -.- is for $A=\frac{\lambda}{V^{2}},B=V^{2},C=0$ . Each
of the cases is shown\ for $\lambda=1$ and $\lambda=4$.
\end{center}}}%
{\parbox[b]{2.8608in}{\begin{center}
\includegraphics[
height=2.8106in,
width=2.8608in
]%
{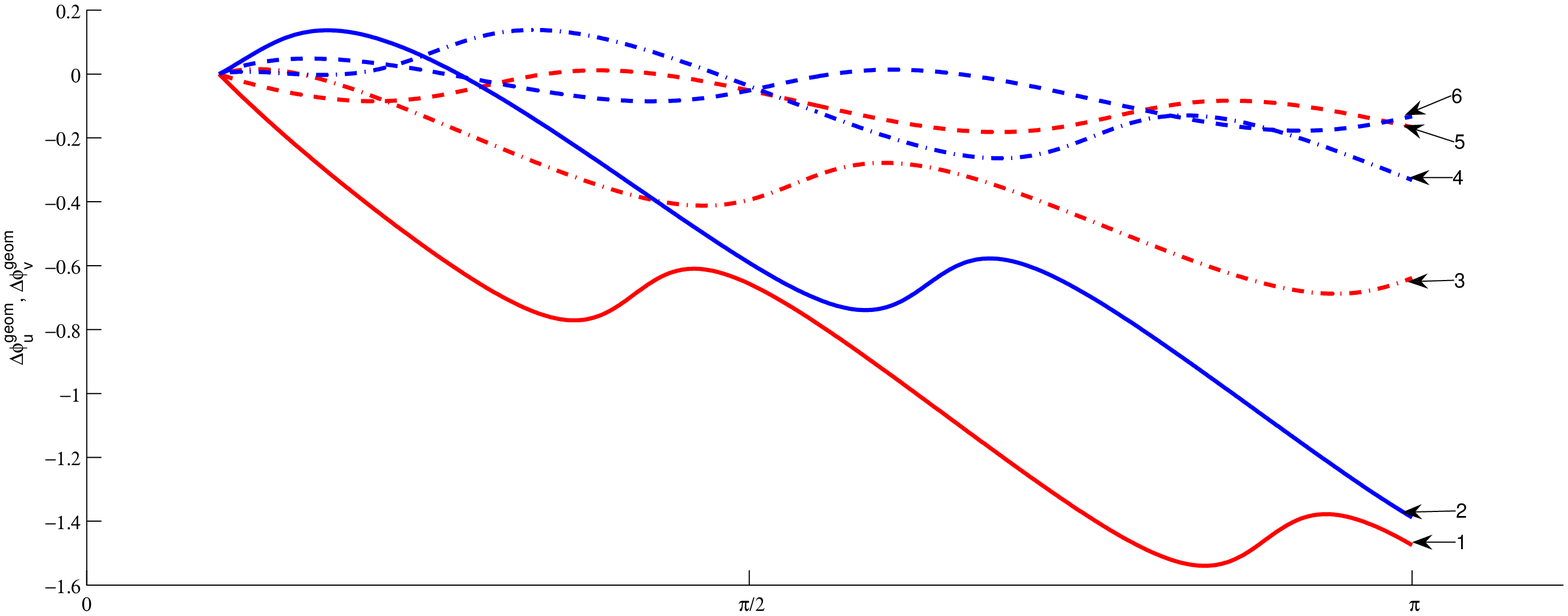}%
\\
Fig. 3. Plots of $\Delta\varphi_{u}^{geom}$ (red) and $\Delta\varphi
_{v}^{geom}$ (blue) given by (\ref{geom}) for the same three cases of the
parameters $A,B$ and $C$ as in Fig.2. Each of the cases is shown\ for
$\lambda=1$.
\end{center}}}%
%

{\parbox[b]{2.8694in}{\begin{center}
\includegraphics[
height=2.7717in,
width=2.8694in
]%
{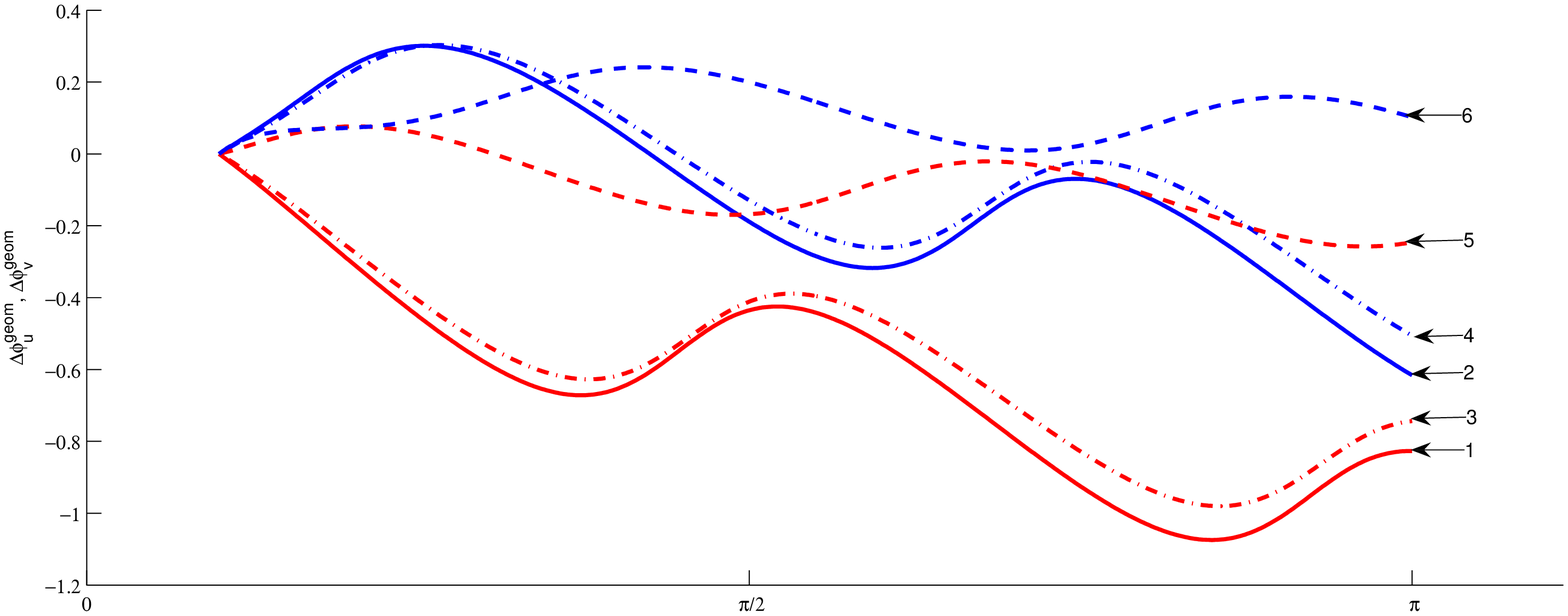}%
\\
Fig. 4. Plots of $\Delta\varphi_{u}^{geom}$ (red) and $\Delta\varphi
_{v}^{geom}$ (blue) given by (\ref{geom}) for the same three cases of the
parameters $A,B$ and $C$ as in Fig.2. Each of the cases is shown for
$\lambda=4$.
\end{center}}}%
{\parbox[b]{2.8608in}{\begin{center}
\includegraphics[
height=2.7622in,
width=2.8608in
]%
{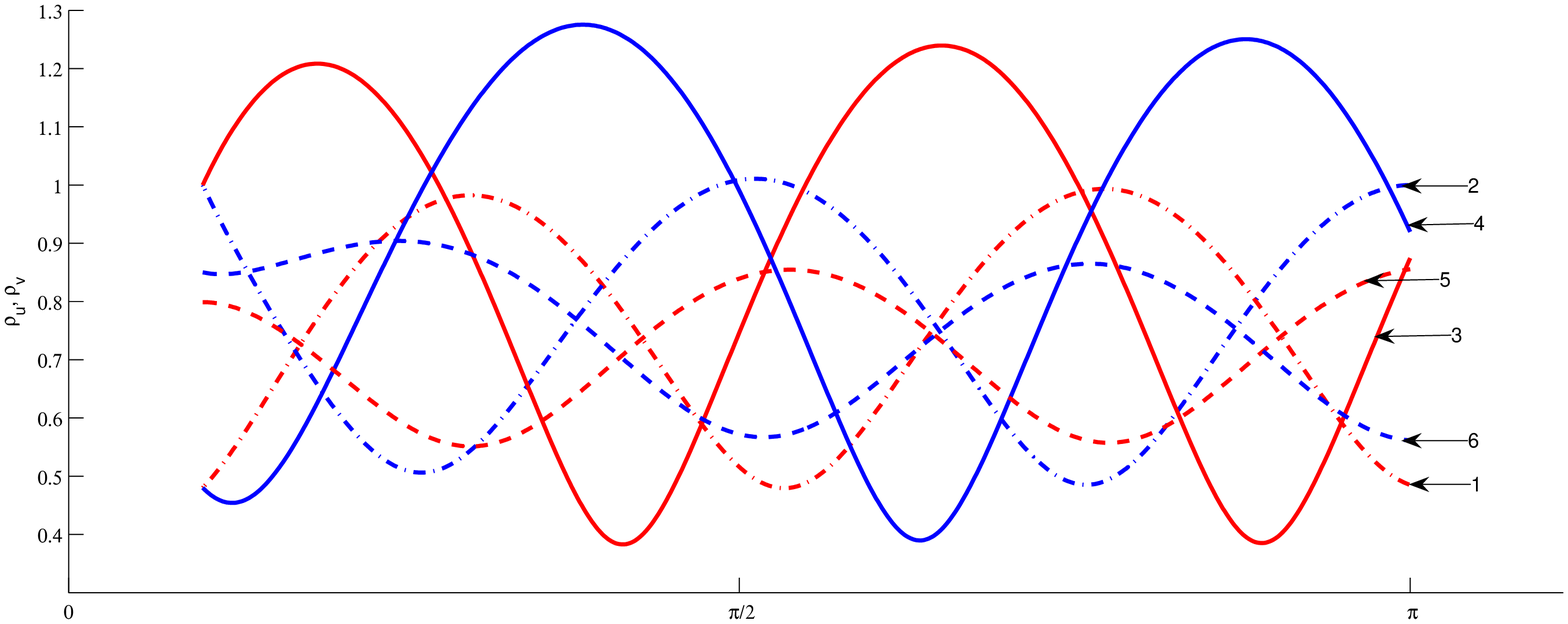}%
\\
Fig.5. The amplitude functions $\rho_{u}$ (red) and $\rho_{v}$ (blue) given by
formulas (\ref{rou}) and (\ref{rov}) for the same parameters $A,B$ and $C$ as
in the previous plots ($\lambda=1).$%
\end{center}}}%

\section{Summary}

After testing Kravchenko's numerical solutions in the context of Jackiw and Pi
model of bilayer graphene we have developed an amplitude-phase approach of
Ermakov-Lewis type for the same model and numerical solutions and for any
radial profile of the gauge potential $\mathcal{A}$. Next, we have chosen a
Debye-screened form of the profile of $\mathcal{A}$ to illustrate our results.
This choice could be of physical relevance when one considers the dynamic
screening processes in the scattering of the charge carriers in the background
of vortex configurations or other topological defects in the bilayer graphene.

Our technique is general and therefore can be applied to any other physical
problem of similar mathematical structure. The geometric phase of
Lewis-Riesenfeld type, which is more general than the Berry phase, is
calculated for the first time in the graphene context. In addition, we
provided a generalization of the Ioffe-Korsch nonlinear intertwining of the
Milne-Pinney amplitude functions.

\section{ Acknowledgments}

The first author would like to thank CONACyT for a postdoctoral fellowship
allowing her to work in IPICyT. The second author wishes to thank CONACyT for
partial support through project 46980. \ The authors thank Dr. R. Jackiw for
useful communications.


\bigskip

\begin{thebibliography}{99}                                                                                               %
\bibitem {nov04}Novoselov K S, Geim A K, Morozov S V, Jiang D, Zhang Y,
Dubonos S V, Grigorieva I V, Firsov A A \ 2004 \textit{Science} \textit{\ }%
\textbf{306} 666

\bibitem {yzhang05}Zhang Y, Tan J W, Stormer H L, Kim P 2005 \textit{Nature}
\textbf{438} 201

\bibitem {mcC-F06}McCann E and Fal'ko V I 2006 \textit{Phys. Rev. Lett.}
\textbf{96} 086805

\bibitem {JP2008}Jackiw R and Pi S-Y 2008 \textit{Phys. Rev.} B \textbf{78} 132104

\bibitem {KrCV2008}Kravchenko V V 2008 \textit{Complex Variables and Elliptic
Equations} \textbf{53} 
775

\bibitem {Hou07}Hou C-Y, Chamon C and Mudry C 2007 \textit{Phys. Rev. Lett.}
\textbf{98} 186809

\bibitem {JP2007}Jackiw R, Pi S-Y\ 2007\textit{\ Phys. Rev. Lett. }\textbf{98
}266402

\bibitem {Milne}Milne E W 1930 \textit{Phys. Rev}. \textbf{35} 863

\bibitem {Pinney}Pinney E 1950 \textit{Proc. Am. Math. Soc. }\textbf{1\ }%
681\textbf{\ }

\bibitem {EliezGr}Eliezer C J and Gray A 1976\ \textit{SIAM J. Appl. Math.}
\textbf{30} 3

\bibitem {Lewis} Lewis H R (Jr) 1967 \textit{Phys. Rev. Lett.} \textbf{18} 510

Lewis H R (Jr) 1968 \textit{J. Math. Phys.} \textbf{9} 1976

\bibitem {Korsch85}Korsch H J 1985 \textit{Phys. Lett. A }\textbf{109} 7

\bibitem {IoffeKorsch}Ioffe M V and Korsch H J 2003 \textit{Phys. Lett.} A
\textbf{311} 200

\bibitem {LR}Lewis H R and Riesenfeld W B 1969 \textit{J. Math. Phys.
}\textbf{10 }1458

\bibitem {mor}Maamache M 1995 \textit{Phys. Rev.} A \textbf{52} 936

Morales D A 1988 \textit{J. Phys.} A \textbf{21} L889

Cerver\'{o} J M and Lejarreta J D 1989 \textit{J. Phys.} A \textbf{22} L663

Espinoza P 2000\textit{\ MS Thesis}, IFUG, Le\'{o}n, Mexico (arXiv: math-ph/0002005)
\end{thebibliography}
\end{document}